\documentstyle[preprint,aps,12pt,epsf]{revtex}
\tightenlines
\begin{document}
\draft
\preprint{\vtop{{\hbox{YITP-02-63}\vskip-0pt
}}}
\title{Newly observed two-body decays of $B$ mesons \\
in a hybrid perspective
}
\author{K. Terasaki\\ Yukawa Institute for Theoretical Physics,
Kyoto University, Kyoto 606-8502, Japan
}
\date{October, 2002}
\maketitle
\thispagestyle{empty}
\begin{abstract}

In consistency with 
$\bar B\rightarrow D^{(*)}\pi$, $J/\psi\bar K$ and $J/\psi\pi$ 
decays, recently observed $B^0\rightarrow D_s^+\pi^-$ and 
%
$\bar B^0\rightarrow D_s^{+}K^-\,$ are
studied in a hybrid perspective in which their amplitude is given 
by a sum of 
%
factorizable and non-factorizable ones.  
\end{abstract}
\vspace{5mm}

(Quasi) two-body decays of $B$ mesons have been studied extensively 
by using the factorization~\cite{fact,Neubert}. However, recently 
measured  rates~\cite{D0-pi0} for the color mismatched  spectator 
(CMS) decays, $\bar B_d^0 \rightarrow D^{(*)0}\pi^0$, are much 
larger than the expectation by the factorization. It suggests that 
non-factorizable contributions can play an important role in 
these decays. In addition, very recently, 
$\bar B^0\rightarrow D_s^+K^-$ and $B^0\rightarrow D_s^+\pi^-$ 
have been observed~\cite{B-Ds-pi}. The rate for the former is 
again much larger than the expectation by the factorization, i.e., 
it has been expected to be strongly suppressed (the helicity 
suppression) since it is described by an annihilation diagram in the 
weak boson mass $m_W\rightarrow\infty$ limit. It means that the 
non-factorizable contribution is dominant in this decay. The latter 
is a pure spectator decay, 
$\bar b\rightarrow \bar u\,+\,(c\bar s)$, 
but does not satisfy the kinematical condition of color 
transparency~\cite{Bjorken}, so that it is not very clear if the 
factorization works well in this decay. Therefore, it is meaningful 
to study a possible role of non-factorizable contributions in the 
newly observed 
$\bar B_d^0 \rightarrow D^{(*)0}\pi^0$, 
$\bar B^0\rightarrow D_s^+K^-$ and $B^0\rightarrow D_s^+\pi^-$ 
in consistency with the $b\rightarrow c$ type of decays, 
$\bar B\rightarrow D^{(*)}\pi$, $J/\psi\bar K$ and $J/\psi\pi$. 

We, first, review briefly our (hybrid) perspective (see 
Ref.\cite{NOVA} for more details). Our starting point is to assume 
that the amplitude can be decomposed into a sum of factorizable and 
non-factorizable ones, ($M_{\rm FA}$ and $M_{\rm NF}$, respectively).
$M_{\rm FA}$ is  estimated by using the factorization while $M_{\rm
NF}$ is assumed to be dominated by dynamical contributions of various
hadron states and calculated by using a hard pion (or kaon)
approximation in the infinite momentum frame
(IMF)~\cite{hard-pion,suppl}. In this approximation, $M_{\rm NF}$ is
given by a sum of the surface term ($M_{\rm S}$) which is given by 
a sum of all possible pole amplitudes and the equal-time commutator 
term ($M_{\rm ETC}$) which arises from the contribution of 
non-resonant (multi-hadron) intermediate states~\cite{Pandit}. 
Corresponding to the above decomposition of the amplitude, 
the effective weak Hamiltonian,  
$H_w\simeq ({G_F/\sqrt{2}})\{c_1O_1 + c_2O_2\}\,+\,h.c.$, 
(where $c_1$ and $c_2$ are the Wilson coefficients), is decomposed
into a sum of the BSW Hamiltonian~\cite{fact}, $H_w^{\rm (BSW)}$, 
and an extra term, $\tilde H_w$, i.e., 
$H_w\rightarrow 
H_w^{\rm (BSW)} + \tilde H_w$, 
by using the Fierz reshuffling, where $H_w^{\rm (BSW)}$ is given by 
a sum of products of colorless currents and might provide the 
factorizable 
amplitude. However, the ``external'' hadron states which sandwich 
$H_w^{\rm (BSW)}$ might interact sometimes with each other through 
hadron dynamics (like a re-scattering, etc.). In this case, 
corresponding part of the amplitude is non-factorizable and should be 
included in $M_{\rm NF}$, so that the values of the coefficients,
$a_1$ and $a_2$, in $M_{\rm FA}$ arising from $H_w^{\rm (BSW)}$ 
might not be the same as the original 
$a_1^{\rm (BSW)}=c_1 + {c_2/N_c}$ and 
$a_2^{\rm (BSW)}=c_2 +{c_1/N_c}$
in $H_w^{\rm (BSW)}$, where $N_c$ is the color degree of freedom. 
Therefore, we will treat $a_1$ and $a_2$ as adjustable parameters 
later. The extra term $\tilde H_w$ which is given by a color singlet 
sum of colored current products provides non-factorizable amplitudes 
in the present perspective, although, in Ref.~\cite{Neubert}, 
contributions from $\tilde H_w$  have been included in the 
factorized amplitudes by considering the effective colors. 

Explicit expression of factorized and non-factorizable amplitudes 
for the $\bar B\rightarrow D^{(*)}\pi$, $J/\psi\bar K$ and 
$J/\psi\pi$ decays have already been given in Ref.~\cite{NOVA} in 
which $M_{\rm ETC}$ and $M_{\rm S}$ with contributions of low lying 
meson poles are  taken into account. In the same way, we can 
calculate the amplitude for the 
$\bar b\rightarrow \bar u\, +\, (c\bar s)$ 
decays, $B^0\rightarrow D_s^{(*)+}\pi^-$. These amplitudes, however, 
include many parameters, i.e., form factors, decay
constants of heavy mesons, asymptotic matrix  elements of $\tilde
H_w$ (matrix elements of
$\tilde H_w$ taken between single hadron states with infinite
momentum), phases,
$\tilde \delta_I$, of $M_{\rm ETC}^{(I)}(\bar B \rightarrow D\pi)$,
($I={1\over 2}$ and ${3\over 2}$), relative to $M_{\rm S}$ and the
relative phase ($\Delta$) between $M_{\rm FA}$ and $M_{\rm NF}$ 
which has not been considered in our previous studies. 
The other parameters involved are known or can be estimated by 
using related experimental data and asymptotic flavor 
symmetries~\cite{Asymp}.  

To obtain improved values of the above amplitudes, we update values 
of parameters involved. Asymptotic matrix elements of axial charges
are estimated as follows, i.e., 
$|\langle{\rho^0|A_{\pi^+}|\pi^-}\rangle| \simeq 1.0$ from 
$\Gamma(\rho\rightarrow\pi\pi)_{\rm exp} \simeq 150$ 
MeV~\cite{PDG00}.   
Here we take 
$\langle{\rho^0|A_{\pi^+}|\pi^-}\rangle = 1.0$ 
and the other ones can be related to it by using related
asymptotic  flavor symmetries, for example, 
$\sqrt{2}\langle{D^{*+}|A_{\pi^+}|D^0}\rangle 
= -\langle{\rho^0|A_{\pi^+}|\pi^-}\rangle$, etc.,  
as in our previous study~\cite{NOVA}. As the values of the
CKM matrix elements~\cite{CKM} and the decay constants, we take 
$V_{cs} \simeq V_{ud} \simeq 0.98$, $V_{cd}\simeq -0.22$, 
$V_{cb}\simeq 0.040$, $|V_{ub}/V_{cb}|\simeq 0.090$ and  
$f_\pi\simeq 130.7$ MeV, $f_K\simeq 160$ MeV from Ref.~\cite{PDG00}. 
The decay constant, $f_{J/\psi}\simeq 406$ MeV, can be obtained from 
$\Gamma(J/\psi\rightarrow e^+e^-)_{\rm exp}=5.26\pm 0.37$ 
keV~\cite{PDG00}. 
The updated values of the decay constants of heavy mesons, 
$f_D\simeq 0.226$ GeV, $f_{D_s}\simeq 0.250$ GeV and 
$f_B\simeq 0.198$ GeV, 
are taken from the lattice QCD~\cite{lattice-DC}, and 
$f_{D^*}\simeq f_D$ and $f_{D_s^*}\simeq f_{D_s}$ 
are assumed as expected by the heavy quark effective theory 
(HQET)~\cite{HQS}. The form factors, $F_0^{(D\bar B)}(m_\pi^2)$ and 
$A_0^{(D^*\bar B)}(m_\pi^2)$, are estimated by using the HQET and 
the data on the semi-leptonic decays of $B$ mesons~\cite{PDG00} as 
$F_0^{(D\bar B)}(m_\pi^2)\simeq 0.74$ and 
$A_0^{(D^*\bar B)}(m_\pi^2)\simeq 0.65$.  
The form factors, 
$F_0^{(\pi\bar B)}(q^2)$ and $F_1^{(\pi\bar B)}(q^2)$, 
are estimated by using extrapolation formulas based on the lattice 
QCD~\cite{lattice-FF}. We here take 
$F_0^{(\pi B)}(m_D^2)\simeq 0.28$, 
$F_0^{(\pi B)}(m_{D_s}^2)\simeq 0.32$, 
$F_1^{(\pi B)}(m_{D^*}^2)\simeq 0.34$, 
$F_1^{(\pi B)}(m_{\psi}^2)\simeq 0.50$ and 
$F_1^{(K B)}(m_{\psi}^2)\simeq 0.59$. 
The annihilation amplitudes which contain $F_0^{(D\pi)}(m_B^2)$ and 
$A_0^{(D^*\pi)}(m_{B}^2)$ will be small and neglected because of the 
helicity  suppression. 

The asymptotic matrix element of $\tilde H_w$ is parameterized by  
\begin{equation}
{\langle {D^{0}|\tilde H_w^{(ud;cb)}|\bar B_d^{0}} \rangle
\over V_{cb}V_{ud}f_\pi} = B_H\times 10^{-5}\,\,{\rm (GeV)}, 
                                                     \label{eq:B_H}
\end{equation}
where $\tilde H_w^{(ud;cb)}$ is a component of $\tilde H_w$ which is 
given by a sum of 
$\tilde O_1^{(ud;cb)}
=V_{ud}V_{cb}\{2\sum_a(\bar dt^au)_L(\bar ct^ab)_L\}$ and 
$\tilde O_2^{(ud;cb)}
=V_{ud}V_{cb}\{2\sum_a(\bar ct^au)_L(\bar dt^ab)_L\}$ 
with the color $SU_c(3)$ generator $t^a$. To evaluate the 
$\bar B\rightarrow D^*\pi$ amplitudes, we assume
\newpage
\begin{quote}
{Table~1. Factorized and non-factorizable amplitudes for the 
$\bar B \rightarrow D^{(*)}\pi$, $J/\psi\bar K$, $J/\psi\pi$ and 
$B^0 \rightarrow D_s^{(*)+}\pi^-$ decays. The CKM matrix elements 
are factored out.}
\end{quote}

\begin{center}
\begin{tabular}
{l c c}
\hline\hline
\vspace{-4mm}
&
&\\
$\quad\,\,${\rm Decay}
&
$\quad A_{\rm FA}\,(\times 10^{-5}$  GeV)
&
$A_{\rm NF}\,(\times 10^{-5}$ GeV) 
\\
&\vspace{-4mm}
\\
\hline 
&\vspace{-3.5mm}
&
\\
$\bar B^0 \rightarrow D^{+}\pi^-$
& $\displaystyle{\quad 1.94\,a_1
e^{i\Delta} }$
&$\displaystyle{- \Bigl\{{4\over 3}e^{i\tilde\delta_{1/2}} 
           -{1\over 3}e^{i\tilde\delta_{3/2}}\Bigr\}B_H}$
\\
& \vspace{-3.5mm}
&\\
\hline 
 & \vspace{-4mm}
&\\
$\bar B^0 \rightarrow D^{0}\pi^0$
& $\displaystyle{\, -1.14\,\Bigl({f_D \over 0.226\,\,{\rm GeV}}\Bigr)
a_2e^{i\Delta}}$
&$\displaystyle{-\Bigl\{{2\sqrt{2} \over 3}e^{i\tilde\delta_{1/2}} 
                  + {\sqrt{2} \over 3}e^{i\tilde\delta_{3/2}}
                                              \Bigr\}B_H}$
\\
 & \vspace{-3.5mm}
&\\
\hline 
 & \vspace{-3.5mm}
&\\
$B^- \rightarrow D^0\pi^-$
& $\displaystyle{\quad 1.94\,a_1\Bigl\{
            1 + 0.48\Bigl({f_D \over f_\pi}\Bigr)
\Bigl({a_2 \over a_1}\Bigr)\Bigr\}e^{i\Delta}}$
& $\displaystyle{\quad e^{i\tilde\delta_{3/2}}B_H}$
\\
 & \vspace{-3.5mm}
&\\
\hline 
 & \vspace{-3.5mm}
&\\
$\bar B^0 \rightarrow D^{*+}\pi^-$
& $\displaystyle{ -1.68\,a_1
e^{i\Delta}}$
& $\displaystyle{ - 0.694 B_H}$
\\
 & \vspace{-4mm}
&\\
\hline 
 & \vspace{-4mm}
&\\
$\bar B^0 \rightarrow D^{*0}\pi^0$
& $\displaystyle{
 \quad 1.07\,\Bigl({f_{D^*}\over 0.226\,\,{\rm GeV}}\Bigr)
a_2e^{i\Delta}}$
&$\displaystyle{\quad0.983B_H}$
\\
& \vspace{-3.5mm}
&\\
\hline 
 & \vspace{-3.5mm}
&\\
$B^- \rightarrow D^{*0}\pi^-$
& $\displaystyle{\, -1.68\,a_1\Bigl\{
         1 + 0.52\Bigl({f_{D^*}\over f_\pi}\Bigr)
\Bigl({a_2 \over a_1}\Bigr)\Bigr\}e^{i\Delta}}$
&   $\displaystyle{- 0.696B_H}$
\\
& \vspace{-3.5mm}
&\\
\hline 
 & \vspace{-3.5mm}
&\\
$B^- \rightarrow J/\psi K^-$
& $\displaystyle{
        -3.60a_2
e^{i\Delta}}$
&$\displaystyle{ -0.548B_H}$
\\
& \vspace{-4mm}
&\\
\hline 
 & \vspace{-3.5mm}
&\\
$\bar B^0 \rightarrow J/\psi\bar K^0$
& $\displaystyle{
        -3.60a_2
e^{i\Delta} }$
& $\displaystyle{-0.548B_H}$
\\
& \vspace{-4mm}
&\\
\hline 
 & \vspace{-3.5mm}
&\\
$B^- \rightarrow J/\psi\pi^-$
& $\displaystyle{ 
       -3.08a_2
e^{i\Delta} }$
 &$\displaystyle{-0.692B_H}$
\\
& \vspace{-4mm}
&\\
\hline 
 & \vspace{-3.5mm}
&\\
$\bar B^0 \rightarrow J/\psi\pi^0$
& $\displaystyle{\quad 
2.18a_2
e^{i\Delta}}$
&$\displaystyle{\quad 0.489B_H
}$
\\
& \vspace{-4mm}
&\\
\hline 
 & \vspace{-3.5mm}
&\\
$B^0 \rightarrow D_s^+\pi^-$
& $\displaystyle{\quad 1.95a_1e^{i\Delta}}$
&$\displaystyle{\quad e^{i\tilde \delta_1}B_H}$
\vspace{-4mm}
\\
 & 
&\\
\hline
 & \vspace{-3.5mm}\\
$B^0 \rightarrow D_s^{*+}\pi^-$
& $\displaystyle{\quad 1.54a_1e^{i\Delta}}$
&$\displaystyle{\quad 0.70B_H}$
\vspace{-4mm}
\\
 & 
&\\
\hline\hline
\end{tabular}

\end{center}
\vspace{5mm}
\begin{equation}
\langle {D^{*0}|\tilde H_w^{(ud;cb)}|\bar B_d^{*0}} \rangle 
= \langle {D^{0}|\tilde H_w^{(ud;cb)}|\bar B_d^{0}} \rangle 
                                           \label{eq:AME-HQS}
\end{equation}
as expected by the HQET. All the other asymptotic 
matrix elements of $\tilde H_w$ involved in the non-factorizable 
amplitudes are combined with the ones in Eq.(\ref{eq:AME-HQS}), 
i.e.,  
\begin{eqnarray}
&&{\langle J/\psi|\tilde H_w^{(cd;cb)}|\bar B_d^{*0} \rangle}
=\Bigl({V_{cd}\over V_{cs}}\Bigr)
{\langle J/\psi|\tilde H_w^{(cs;cb)}|\bar B_s^{*0}\rangle}
\nonumber\\
&&=\Bigl({V_{cd}\over V_{ud}}\Bigr)
   {\langle{D^{(*)0}|\tilde H_w^{(ud;cb)}|\bar B_d^{(*)0}}\rangle} 
\nonumber\\
&&=-\Bigl({V_{cd}V_{cb}\over V_{cs}V_{ub}}\Bigr)
      {\langle{D_s^{(*)+}|\tilde H_w^{(cs;ub)}|B_u^{(*)+}}\rangle},
\label{eq:MEH}
\end{eqnarray}
by inserting commutation relations, 
$[V_{K^0}, \tilde H_w^{(cs;cb)}] 
= ({V_{cs}/ V_{cd}})\tilde H_w^{(cd;cb)}$, 
$[V_{D^0}, \tilde H_w^{(cd;cb)}] 
= ({V_{cd}/ V_{ud}})\tilde H_w^{(ud;cb)}$, 
$[V_{\bar D^0}, \tilde H_w^{(cs;cb)}] 
= ({V_{cb}/ V_{ub}})\tilde H_w^{(cs;ub)}$,   
between related asymptotic states (single hadron states with infinite 
momentum) and using asymptotic $SU_f(3)$ and $SU_f(4)$ relations, 
${\langle {\bar B_s^{*0}|V_{K^0}|\bar B_d^{*0}} \rangle }= -1$, 
${\langle {D^{*0}|V_{D^0}|J/\psi} \rangle} = -1$, etc. 
To obtain the last equality in Eq.(\ref{eq:MEH}), we have used the 
$CP$-invariance which is always assumed in this note and 
$\langle{\{q\bar q\}_0|\tilde O_+|\{q\bar q\}_0}\rangle = 0$ 
from a quark counting~\cite{note}, where 
$\tilde O_\pm = \tilde O_1\pm \tilde O_2$.  
The $\{q\bar q\}_0$'s denote the low lying mesons. 
\newpage
\begin{quote}
{Table~2. A typical result on the branching ratios 
($\times 10^{-3}$) for $\bar B \rightarrow D^{(*)}\pi$, 
$J/\psi\bar K$ and $J/\psi\pi$ decays, where the values of the
parameters involved are given in the text. 
${\cal B}_{\rm FA}$ and ${\cal B}_{\rm tot}$ are given 
by ${M}_{\rm FA}$ and ${M}_{\rm tot}$, respectively. 
${\cal B}_{\rm exp}$ are taken from Ref.~\cite{PDG02}.} 
\end{quote}

\begin{center}
\begin{tabular}
{l l l c}
\hline\hline
\vspace{-4mm}
&
&\\
 {\quad Decays} 
&\quad ${\cal B}_{\rm FA}$ \quad
&\quad ${\cal B}_{\rm tot}$ \quad
& ${\cal B}_{\rm exp}$  \qquad
\\
\vspace{-4mm}
&
&\\
\hline
\vspace{-4mm}
&
&\\
{$\bar B^0\rightarrow D^+\pi^-$}
&
\hspace{2mm}{4.0}
&
\hspace{2mm}{3.1}
&$3.0 \pm 0.4$
\\
\vspace{-4mm}
&
&\\
\hline
\vspace{-4mm}
&
&\\
{$\bar B^0\rightarrow D^0\pi^0$}
&
\hspace{2mm}{0.10}
&
\hspace{2mm}{0.24}
&$0.27\pm 0.06$
\\
\vspace{-4mm}
&
&\\
\hline 
\vspace{-4mm}
&
&\\
{$B^-\rightarrow D^0\pi^-$}
&
\hspace{2mm}{5.6}
&
\hspace{2mm}{5.6}
&$5.3 \pm 0.5$
\\
\vspace{-4mm}
&
&\\
\hline
\vspace{-4mm}
&
&\\
{$\bar B^0\rightarrow D^{*+}\pi^-$}
&
\hspace{2mm}{3.1}
&
\hspace{2mm}{2.6}
&$2.76 \pm 0.21$
\\
\vspace{-4mm}
&
&\\
\hline
\vspace{-4mm}
&
&\\
{$\bar B^0\rightarrow D^{*0}\pi^0$}
&
\hspace{2mm}{0.09}
&
\hspace{2mm}{0.22}
&$0.22\pm 0.10$
\\
\vspace{-4mm}
&
&\\
\hline
\vspace{-4mm}
&
&\\
{$\bar B^0\rightarrow D^{*0}\pi^-$}
&
\hspace{2mm}{4.1}
&
\hspace{2mm}{4.7}
&$4.6 \pm 0.4$
\\
\vspace{-4mm}
&
&\\
\hline
\vspace{-4mm}
&
&\\
{$B^- \rightarrow J/\psi K^-$}
&
\hspace{2mm}{0.82}
&
\hspace{2mm}{0.99}
&$1.01 \pm 0.05$
\\
\vspace{-4mm}
&
&\\
\hline
\vspace{-4mm}
&
&\\
{$\bar B^0\rightarrow J/\psi\bar K^0$}
&
\hspace{2mm}{0.75}
&
\hspace{2mm}{0.91}
&$0.87 \pm 0.05$
\\
\vspace{-4mm}
&
&\\
\hline
\vspace{-4mm}
&
&\\
{$B^-\rightarrow J/\psi\pi^-$}
&
\hspace{2mm}{0.030}
&
\hspace{2mm}{0.039}
&\hspace{2mm}$0.042 \pm 0.007$\hspace{2mm}
\\
\vspace{-4mm}
&
&\\
\hline
\vspace{-4mm}
&
&\\
{$\bar B^0\rightarrow J/\psi\pi^0$}
&
\hspace{2mm}{0.014}
&
\hspace{2mm}{0.018}
&$0.021\pm 0.005$
\\
\hline\hline
\end{tabular}
\end{center}
\vspace{3mm}
In this way, we  can obtain $M_{\rm FA}$ and $M_{\rm NF}$ in the 
second and third columns, respectively, of Table~1, where we have 
neglected small contributions of annihilation terms in $M_{\rm FA}$ 
and excited meson poles in $M_{\rm NF}$. 

We now look for values of parameters, $a_1$, $a_2$, $\Delta$, 
$\tilde\delta_{I}$, ($I={1\over 2}$ and ${3\over 2}$), and $B_H$, 
which reproduce the measured branching ratios for the 
$\bar B\rightarrow D^{(*)}\pi$, $J/\psi\bar K$ and $J/\psi\pi$ 
decays. $a_1$ and $a_2$ are treated as adjustable parameters 
with values around $a_1^{\rm (BSW)}$ and $a_2^{\rm (BSW)}$. 
The phase $\tilde\delta_{I}$ is restricted in the region 
$|\tilde\delta_{I}| < 90^\circ$ since resonant contributions have 
already been extracted as pole amplitudes in $M_{\rm S}$ while 
$\Delta$ and $B_H$ are treated as free parameters. The result is not 
very sensitive to $\tilde\delta_{I}$, and the coefficients, $a_1$ 
and $a_2$, favor values close to the ones taken in 
Ref.~\cite{Neubert} which is based on the factorization. 
The above implies that the non-factorizable contribution is not very 
important in the color  favored decays.  We can reproduce the 
experimental data  (${\cal B}_{\rm exp}$) compiled by the 
Particle~Data~Group~2002~\cite{PDG02} taking values of 
parameters in the range, 
$1.00\lesssim a_1 \lesssim 1.13$, $0.28\lesssim a_2\lesssim 0.31$, 
$24^\circ\lesssim |\Delta| \lesssim 32^\circ$,  
$|\delta_{1/2}|\lesssim 70^\circ$, 
$10^\circ\lesssim |\delta_{3/2}|\lesssim 90^\circ$ 
and $0.09\lesssim B_H \lesssim 0.25$. 
To see more explicitly a role of the non-factorizable contribution, 
we list a typical result on the branching ratios (near the best fit 
to ${\cal B}_{\rm exp}$) for 
$a_1=1.08$, $a_2=0.29$, $\tilde\delta_1=0.0^\circ$, 
$\tilde\delta_3=\pm 90^\circ$, $\Delta=\pm 28^\circ$ and 
$B_H=0.19$ 
in Table~2, where we have used 
$\tau(B^-) = 1.67\times10^{-12}$~s and 
$\tau(\bar B^0) = 1.54\times10^{-12}$~s from Ref.~\cite{PDG02}. 
${\cal B}_{\rm FA}$ and ${\cal B}_{\rm tot}$ are given by 
$M_{\rm FA}$ and $M_{\rm tot}=M_{\rm FA}\,+\,M_{\rm NF}$,
respectively. As seen in Table 2, ${\cal B}_{\rm FA}$ in which 
$M_{\rm NF}$ is discarded is hard to reproduce the data on the CMS 
decays, $\bar B\rightarrow D^{(*)0}\pi^0$. If we add $M_{\rm NF}$, 
however, we can get a much better fit to the data including the CMS 
decays. In the color favored $\bar B\rightarrow D^{(*)}\pi$ decays, 
$M_{\rm NF}$ is rather small (but it can interfere efficiently with 
the main amplitude, $M_{\rm FA}$). In the $B^-\rightarrow D^0\pi^-$ 
decay, however, it is very small. 
In the $\bar B\rightarrow J/\psi\bar K$  and $J/\psi\pi$ decays, the
color suppression does not work so well that $M_{\rm NF}$ is not 
dominant in contrast with the $\bar B\rightarrow D^{(*)0}\pi^0$ 
although all of them  are the CMS decays. 

Next, we study the $B^0\rightarrow D_s^{(*)+}\pi^-$ decays comparing 
with the $B^-\rightarrow D^0\pi^-$ which has been studied above. 
Using the same values of parameters as the above, i.e., $a_1=1.08$, 
$a_2=0.29$, $B_H=0.19$, we obtain 
\begin{eqnarray}
&&|{M_{\rm NF}(B^0\rightarrow D_s^+\pi^-)|
         \simeq 0.09|M_{\rm FA}(B^0\rightarrow D_s^+\pi^-)}|, \\
&&|{M_{\rm NF}(B^0\rightarrow D_s^{*+}\pi^-)|        
\simeq 0.08|M_{\rm FA}(B^0\rightarrow D_s^{*+}\pi^-)}|, 
\end{eqnarray}
which imply that the factorization works considerably well in these 
decays although they do not satisfy the condition of the color 
transparency. Neglecting the rather small $M_{\rm NF}$ in the 
$B^0\rightarrow D_s^+\pi^-$ and using the same values of parameters 
as the above, we obtain 
\begin{equation}
|{M(B^0\rightarrow D_s^+\pi^-)|        
\simeq 0.074|M(B^-\rightarrow D^0\pi^-)}| , 
\end{equation}
where we have used 
$|V_{ub}/V_{cb}|_{\rm exp}\simeq 0.090$~\cite{PDG00}.
The measured branching ratio for the $B^-\rightarrow D^0\pi^-$ 
decay~\cite{PDG02} leads us to 
\begin{equation}
{\cal B}(B^0\rightarrow D_s^+\pi^-)\simeq 2.7\times 10^{-5},
\end{equation}
which reproduces well the recent measurements~\cite{B-Ds-pi}, 
\[\left\{
\begin{array}{l}
\displaystyle{
{\cal B}(B^0\rightarrow D_s^+\pi^-)_{\rm BABAR}
                               = (3.1 \pm 2.0)\times 10^{-5}},\\
\displaystyle{
{\cal B}(B^0\rightarrow D_s^+\pi^-)_{\rm BELLE}
            = (2.4 ^{+1.0}_{-0.8}\pm 0.7)\times 10^{-5}}.
\end{array}
\right.
\]
In the same way, we obtain 
${\cal B}(B^0\rightarrow D_s^{*+}\pi^-)\simeq 1.7\times 10^{-5}$, 
which is again compatible with the experimental 
upper limits~\cite{B-Ds-pi}. 

In the $\bar B^0 \rightarrow D_s^+K^-$ decay, $M_{\rm FA}$ is 
strongly suppressed because of the helicity suppression, so that 
$M_{\rm NF}$ dominates the decay in the present perspective, i.e., 
\begin{eqnarray}
&&M(\bar B^0\rightarrow D_s^+K^-)
\simeq M_{\rm NF}(\bar B^0\rightarrow D_s^+K^-)  \nonumber\\
&&\hspace{15mm}
\simeq -iV_{cb}V_{ud}\Bigl({f_\pi\over f_K}\Bigr)
{\langle{D^0|\tilde H_w|\bar B_d^0}\rangle
\over V_{cb}V_{ud}f_\pi}e^{i\tilde\delta_1}. 
\end{eqnarray}
The same value of parameters as the above leads to 
\begin{equation}
{\cal B}(\bar B^0\rightarrow D_s^+K^-)
\simeq 2.8\times10^{-5},
\end{equation}
which should be compared with the measured values~\cite{B-Ds-pi}, 
\[\left\{
\begin{array}{l}
\displaystyle{
{\cal B}(\bar B^0\rightarrow D_s^+K^-)_{\rm BABAR}
                               = (3.2 \pm 2.0)\times 10^{-5}},\\
\displaystyle{
{\cal B}(\bar B^0\rightarrow D_s^+K^-)_{\rm BELLE}
            = (4.6 ^{+1.2}_{-1.1}\pm 1.3)\times 10^{-5}}.
\end{array}
\right.
\]

In summary, we have studied the recently observed decays, 
$\bar B\rightarrow D^{(*)0}\pi^0$, $B^0 \rightarrow D_s^{+}\pi^-$ 
and $\bar B^0 \rightarrow D_s^+K^-$, 
in consistency with the $b\rightarrow c$ type of decays, 
$\bar B \rightarrow D^{(*)}\pi$, $J/\psi\bar K$ and $J/\psi\pi$, 
providing their amplitude by a sum of factorized and 
non-factorizable ones. To study the non-factorizable 
amplitudes, we have used the asymptotic $SU_f(3)$ and $SU_f(4)$ 
symmetries which may be broken. The size of the symmetry breaking 
can be estimated from the value of the form factor, $f_+(0)$, in 
the matrix element of related vector current, where 
$f_+(0)=1$ in the symmetry limit. From the measured values of the 
form factors, 
$f_+^{(\pi D)}(0)=0.71 \pm 0.06$~\cite{E687-1} and 
$|f_+^{(\pi D)}(0)/f_+^{(\bar K D)}(0)|=1.00 \pm 0.13$~\cite{E687-2}, 
the asymptotic $SU_f(4)$ symmetry seems to be broken to the extent 
of 30~$\%$ while the asymptotic $SU_f(3)$ still works well. However, 
such a large symmetry breaking has not caused any serious problem in 
the present study since $M_{\rm NF}$ is much smaller than 
$M_{\rm FA}$ except for some decays in which $M_{\rm FA}$ is 
strongly suppressed and whose experimental errors are still large. 
For more precise studies, of course, more detailed informations of 
the symmetry breaking will be needed. 

The amplitude with final state interactions has been included in 
the non-factorizable one. For the color favored 
$\bar B\rightarrow D^{(*)}\pi$ decays, $M_{\rm NF}$ has been 
rather small and, therefore, the final state interactions seem to 
be not very important (but not necessarily negligible) in these 
decays. In the $\bar B\rightarrow D^{(*)0}\pi^0$ which are the 
CMS decays, $M_{\rm NF}$ has been dominant since $M_{\rm FA}$ is 
suppressed because of the color suppression.  In the 
$\bar B\rightarrow J/\psi\bar K$ and $J/\psi\pi$, which also are 
the CMS decays, however, the color suppression has not worked so 
well that $M_{\rm NF}$ has not been dominant in contrast with the 
$\bar B \rightarrow D^{(*)0}\pi^0$ decays. In the 
$B^0\rightarrow D_s^{(*)+}\pi^-$ which are the color favored
$\bar b\rightarrow \bar u\,+\,(c\bar s)$ type of spectator decays, 
$M_{\rm NF}$ has been small. The values of parameters which 
reproduce the measured  branching ratios for the $\bar B\rightarrow 
D^{(*)}\pi$, $J/\psi\bar K$ and $J/\psi\pi$ decays have lead to 
${\cal B}(B^0\rightarrow D_s^{(*)+}\pi^-)$ consistent with the 
very recent measurements. It means that the factorization works 
considerably well in these decays although they do not satisfy 
the condition of color transparency. In the 
$\bar B^0\rightarrow D_s^+K^-$ which is the annihilation decay, 
$M_{\rm NF}$ has been dominant and reproduced the very recent 
measurements within their large errors. All the above suggest 
that dynamical contributions of hadrons should  be carefully 
treated in hadronic  weak decays of $B$ mesons. 

In the CMS decays, 
$\bar B^0 \rightarrow  D^{(*)0}\pi^0$, $J/\psi\bar K$ and 
$J/\psi\pi$, the annihilation decay, $\bar B^0\rightarrow D_s^+K^-$, 
and the $\bar b\rightarrow \bar u\, +\, (c\bar s)$ type of 
spectator decay, $B^0\rightarrow D_s^+\pi^-$, 
both of the theoretical and experimental ambiguities are still 
large although their measured rates have been reproduced considerably 
well by taking account of the non-factorizable contributions. 
More theoretical and experimental studies on these decays will 
be needed. 

\vspace{5mm}
\thanks
{The author would like to appreciate Prof. T.~Onogi for discussions. 
He also thanks the Yukawa Institute for Theoretical Physics at Kyoto
University. Discussions during the YITP workshop, YITP-W-02-08, 
``TEA (Theoretical-Experimental-Astronomical Particle Physics) 02'',  
were useful to complete this work.}


\begin{references}
\bibitem{Asymp} Asymptotic flavor symmetry is the flavor symmetry of 
asymptotic matrix elements -- matrix elements of charges, currents 
and some other operators (like the weak Hamiltonian constructed by 
them) taken between single hadron states with infinite momentum. 
A comprehensive review on the asymptotic flavor symmetry and its 
fruitful results has been provided in Ref.\cite{suppl}.
\end{references}
\end{document}